\documentclass[conference]{IEEEtran}
\IEEEoverridecommandlockouts
\usepackage{amsmath}
\usepackage{graphicx}
\usepackage{url}
\usepackage{hyperref}
\usepackage{cite}
\usepackage{amsfonts}
\usepackage{mathtools}
\usepackage{hhline}
\usepackage[caption=false, font=footnotesize]{subfig}

\usepackage{algpseudocode}
\usepackage{algorithm}
\algnewcommand\algorithmicforeach{\textbf{for each}}
\algdef{S}[FOR]{ForEach}[1]{\algorithmicforeach\ #1\ \algorithmicdo}

\usepackage[table]{xcolor}
\usepackage{commath,multirow}
\usepackage{tabularx} 
\usepackage{diagbox}

\usepackage{verbatim} 

\usepackage{adjustbox,cleveref}

\newcommand{\etal}{\textit{et al}. }

\makeatletter
\def\ps@IEEEtitlepagestyle{%
  \def\@oddfoot{\mycopyrightnotice}%
  \def\@evenfoot{}%
}
\def\mycopyrightnotice{
  {\footnotesize
  \begin{minipage}{\textwidth}
  \centering
  Copyright~\copyright~2021 IEEE. Personal use of this material is permitted. Permission from IEEE must be obtained for all other uses, in any current or future media, including reprinting/republishing this material for advertising or promotional purposes, creating new collective works, for resale or redistribution to servers or lists, or reuse of any copyrighted component of this work in other works
  \end{minipage}
  }
  \gdef\mycopyrightnotice{}
}

\begin{document}

\title{High Frame Rate Video Quality Assessment using VMAF and Entropic Differences}
\author{Pavan C Madhusudana$^1$, \and Neil Birkbeck$^2$, \and Yilin Wang$^2$, \and Balu Adsumilli$^2$, \and Alan C. Bovik$^1$ \and
\centerline{$^1$The University of Texas at Austin, TX, USA. $^2$Google, Mountain View, CA, USA} \and \centerline{E-mail : pavancm@utexas.edu, \{birkbeck, yilin, badsumilli\}@google.com, bovik@ece.utexas.edu}}
    

\maketitle

\begin{abstract}
The popularity of streaming videos with live, high-action content has led to an increased interest in High Frame Rate (HFR) videos. In this work we address the problem of frame rate dependent Video Quality Assessment (VQA) when the videos to be compared have different frame rate and compression factor. The current VQA models such as VMAF have superior correlation with perceptual judgments when videos to be compared have same frame rates and contain conventional distortions such as compression, scaling etc. However this framework requires additional pre-processing step when videos with different frame rates need to be compared, which can potentially limit its overall performance. Recently, Generalized Entropic Difference (GREED) VQA model was proposed to account for artifacts that arise due to changes in frame rate, and showed superior performance on the LIVE-YT-HFR database which contains frame rate dependent artifacts such as judder, strobing etc. In this paper we propose a simple extension, where the features from VMAF and GREED are fused in order to exploit the advantages of both models. We show through various experiments that the proposed fusion framework results in more efficient features for predicting frame rate dependent video quality. We also evaluate the fused feature set on standard non-HFR VQA databases and obtain superior performance than both GREED and VMAF, indicating the combined feature set captures complimentary perceptual quality information. 
\end{abstract}


\section{Introduction}
The explosion of internet usage has resulted in videos becoming more mainstream and forming an important component in day-to-day lives. With videos commanding a significant share in internet traffic, considerable effort has been made in improving the video characteristics in terms of dynamic range, spatial resolutions (4K/8K), immersive content (Virtual/Augmented Reality) and extending temporal resolutions by employing higher frame rate (HFR) videos. However, there has been much less emphasis placed on increasing frame rates, with many existing display devices and content compatible only upto 60 frames per second (fps).

With the recent surge in streaming content pertaining to sports, high motion and live action, there has been a renewed interest in HFR videos. It is a common belief that using more number of frames results in smoother variation of motion content, thus higher frame rates results in better quality videos. However, there have been very few works which comprehensively evaluate these notions. Existing legacy Video Quality Assessment (VQA) databases such as LIVE-VQA \cite{seshadrinathan2010study}, CSIQ-VQA \cite{vu2014vis3} only contain videos which have reference and distorted versions at the same frame rate. This restricts the analysis of distortions that arise due to frame rate changes. Recent publications of HFR-VQA databases such as Waterloo-HFR \cite{nasiri2015perceptual}, BVI-HFR \cite{mackin2018study} and LIVE-YT-HFR \cite{pavan2020liveythfr} has rekindled interest in the problem of HFR-VQA.

Perceptual VQA models are an essential component of many video streaming applications such as YouTube, Netflix, Hulu etc. where video quality is objectively evaluated at a large scale. VQA models can be categorized into three divisions: Full Reference (FR) models which have access to pristine high quality reference \cite{wang2004image,wang2003multiscale,sheikh2006image,zhang2011fsim,seshadrinathan2009motion}, Reduced Reference (RR) models which use limited reference information \cite{soundararajan2012rred,soundararajan2012video} and No Reference (NR) models which use only the distorted version for quality prediction \cite{mittal2012no,mittal2013making,saad2014blind}. In this paper we consider the problem of VQA when reference and distorted sequences can possibly have different frame rates, thus it falls under the category of FR and RR VQA methods.

Prior works addressing the problem of VQA in HFR domain are very limited. In \cite{nasiri2018temporal} a motion smoothness measure is introduced to quantify quality changes that occur due to frame rate variations. Zhang \etal \cite{zhang2017frame} proposed Frame Rate Quality Metric (FRQM) which uses differences between the wavelet coefficients of reference and distorted videos to measure quality, and achieved competitive performance on the BVI-HFR dataset. Although the above methods account for artifacts arising due to frame rate changes, their performance deteriorates in the presence of compression, thus limiting their applicability. In \cite{madhusudana2020stgreed} a model based on entropic differences in spatial and temporal band-pass domain was proposed to capture frame rate artifacts in the presence of compression, and achieved superior performance on the LIVE-YT-HFR database.

In this work we propose a simple fusion framework to combine features from VMAF \cite{VMAF2016} and GREED \cite{madhusudana2020stgreed} models. This principle is motivated from \cite{bampis2018spatiotemporal} where similar idea was presented to combine VMAF features with SpEED \cite{bampis2017speed} VQA model. VMAF has superior correlation with subjective judgments on existing FR-VQA databases. However VMAF performance is limited when used with videos of different frame rates. Since these videos contain dominant temporal artifacts, lower VMAF performance indicates that the temporal component in VMAF is insufficient to capture frame rate artifacts. On the other hand, GREED achieves good performance when used with videos of different frame rates, suggesting the superiority of temporal features present in GREED. Thus fusing these features can result in a robust model that can perform well on both HFR and non-HFR VQA databases. 

The rest of the paper is organized as follows. In Section \ref{sec:GREED} we provide a brief background about GREED and VMAF, and discuss the proposed feature integration procedure. In Section \ref{sec:experiments} we report and analyze various experimental results, and provide some concluding remarks in Section \ref{sec:conclusion}.

\section{Background}
\label{sec:GREED}

\subsection{GREED Description}
A brief description of GREED \cite{madhusudana2020stgreed} model and its components is provided in this section. GREED model is motivated from the statistical deviations observed between the distributions of band-pass coefficients of different frame rates. The video $V(\mathbf{x},t)$ ($\mathbf{x} = (x,y)$ is subjected to temporal band-pass filtering by a bank of $K$ 1D filters denoted by $b_k$ for $k \in \{1,\ldots K\}$
\begin{align}
    B_k(\mathbf{x},t) = V(\mathbf{x},t)*b_k(t) \text{\hspace{10pt}} \forall k \in \{1,\ldots K\},
\end{align}
where $B_k$ denotes band-pass response of $k^{th}$ filter. The coefficients of $B_k$ show heavy tailed distribution which can be well modeled using Generalized Gaussian Distribution (GGD) which is given by
\begin{align*}
    f(x;\alpha,\beta) = \frac{\beta}{2\alpha\Gamma(1/\beta)}\exp\Big(-\Big(\frac{|x|}{\alpha}\Big)^{\beta}\Big)
\end{align*}
where $\Gamma(.)$ is the gamma function:
\begin{align*}
    \Gamma(a) = \int_0 ^{\infty}x^{a-1}e^{-x} dx .
\end{align*}
Here $\alpha$ and $\beta$ are parameters of GGD. The band-pass coefficients are partitioned into non-overlapping patches of size $\sqrt{M} \times \sqrt{M}$, which are indexed by $p \in \{ 1,2, \ldots P \}$. Let $B_{kpt}$ denote the vector of band-pass coefficients at frame $t$, patch $p$ and band-pass filter $k$. The band-pass coefficients $B_{kpt}$ are allowed to pass through a Gaussian channel to model perceptual imperfections such as neural noise \cite{sheikh2006image,soundararajan2012video} resulting in a noisy version $\Tilde{B}_{kpt}$ which can also be approximately modeled using a GGD. The main idea here is to compare the entropy $h(\Tilde{B}_{kpt}$ of reference and distorted versions for evaluating quality. The expression for GGD entropy of random variable $X \sim GGD(0,\alpha,\beta)$ is given by:
\begin{align}
    h(X) = \frac{1}{\beta} - \log \left(\frac{\beta}{2\alpha\Gamma(1/\beta) }\right).
    \label{eqn:ggd_entropy}
\end{align}
$h(\Tilde{B}_{kpt})$ is calculated using (\ref{eqn:ggd_entropy}) for each patch by estimating parameters $\alpha$ and $\beta$ using kurtosis matching procedure detailed in \cite{soury2015new,pan2012exposing}.

\begin{figure}[t]
    \captionsetup[subfigure]{justification=centering}
        \centering
        \subfloat[Content - \texttt{bouncyball}]{\includegraphics[width=0.48\linewidth]{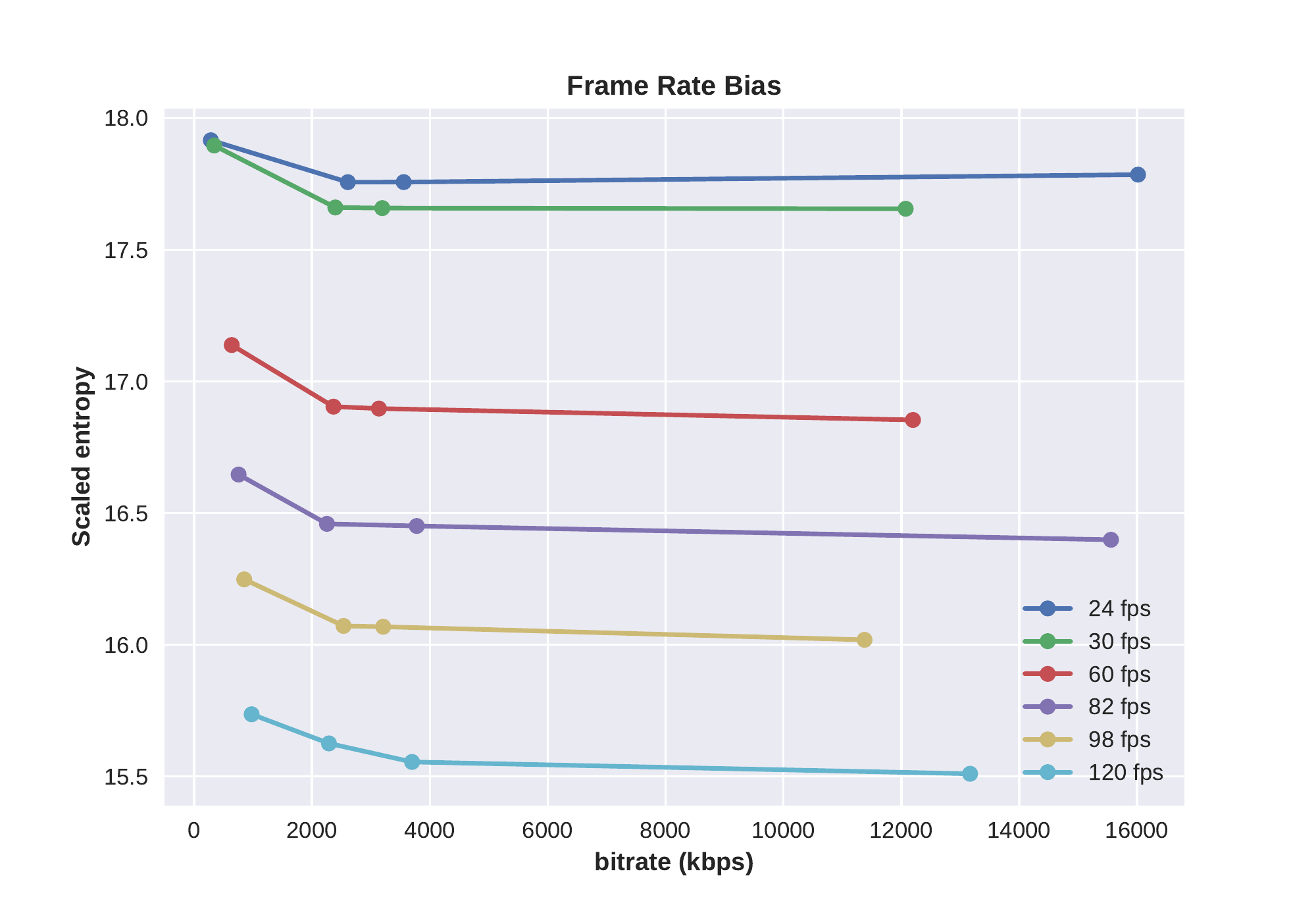}} \quad
        \subfloat[Content - \texttt{leaves}]{\includegraphics[width=0.48\linewidth]{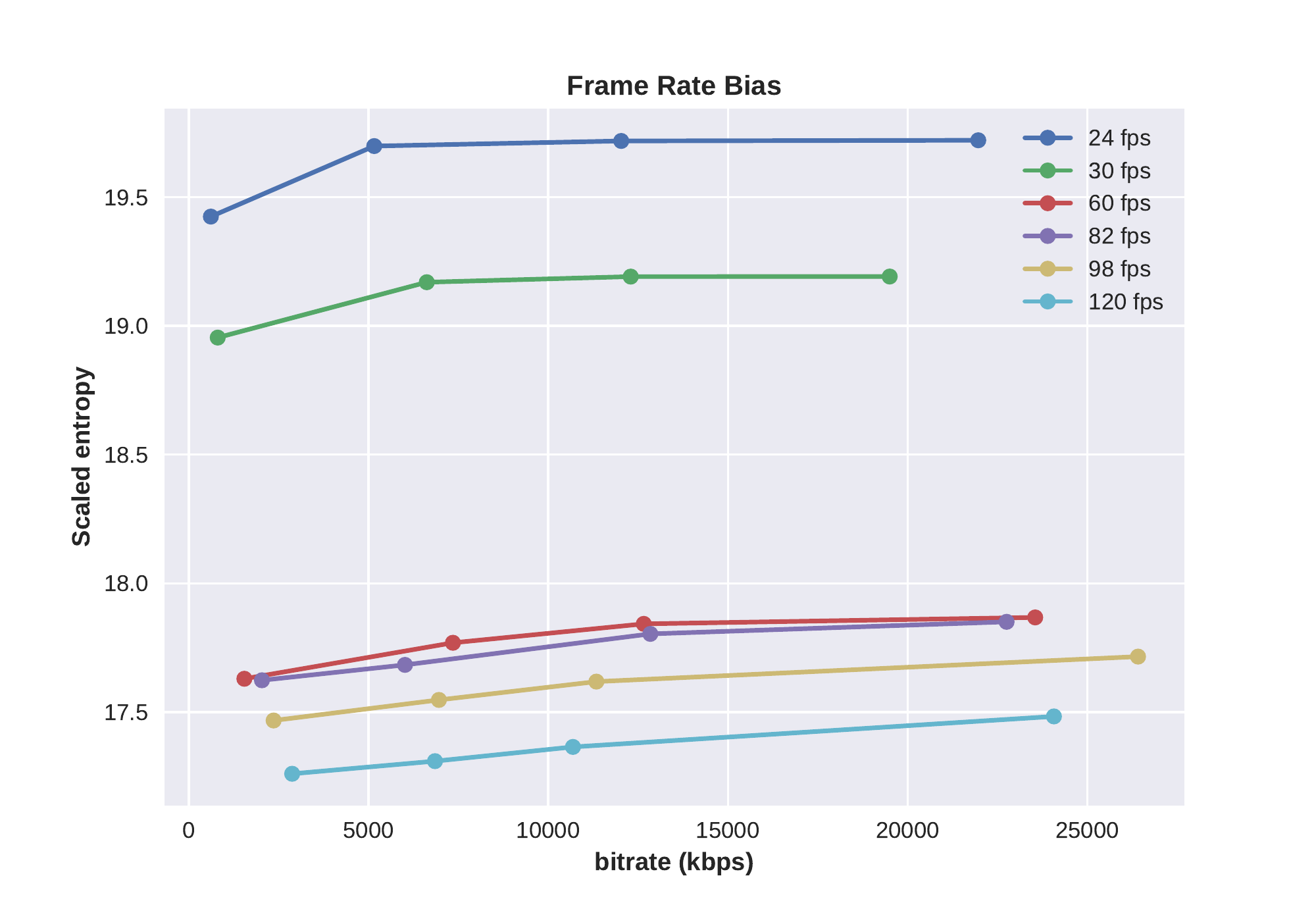}}
        \caption{Plot of scaled entropies against bitrate.}
        \label{fig:frame_rate_bias}
\end{figure}

\paragraph*{\textbf{Temporal GREED}}
To make entropies more local and provide numerical stability to regions with small variances, $h(\Tilde{B}_{kpt})$ is premultiplied by a scaling factor given as:
\begin{align*}
    \epsilon_{kpt} = [\log(1+\sigma^2(\Tilde{B}_{kpt}))]h(\Tilde{B}_{kpt})
\end{align*}
Subsampling a signal as well as compression leads to entropy modification. However the entropy change due to frame rate dominates that occurring due to compression as can be observed from Fig. \ref{fig:frame_rate_bias} where entropy values remain nearly same regardless of compression factor. This is referred to as frame rate bias of entropy and this makes entropy difference inadequate to account for compression distortions. To overcome this, reference video $R$ is temporally downsampled to have the same frame rate of distorted video $D$ resulting in another sequence called Pseudo Reference $PR$. Given these signals, Temporal GREED is defined as
\begin{align}
    \text{TGREED}_{kt} = \frac{1}{P}\sum_{p=1} ^P \Bigg|\Big(1 + |\epsilon_{kpt} ^D - \epsilon_{kpt} ^{PR}|\Big) \frac{\epsilon_{kpt} ^R + 1}{\epsilon_{kpt} ^{PR} + 1} - 1 \Bigg|
    \label{eqn:GTI}
\end{align}

TGREED can be interpreted as a weighted entropic difference, where the ratio term which is dependent on frame rate of $R$ and $D$ acts as a weighting factor. From Fig. \ref{fig:frame_rate_bias} it can be observed that the entropy separation among various frame rates is dependent on the content. This makes the ratio term sensitive to content. The absolute difference term captures compression artifacts since $PR$ is lossless version of $D$ at the same frame rate.

\paragraph*{\textbf{Spatial GREED}}: For capturing spatial distortions, each frame is subjected to spatial band-pass filtering by local Mean Subtracted (MS) filtering scheme similar to \cite{bampis2017speed}. Similar to temporal band-pass coefficients, MS coefficients also can be modeled using GGD and scaled entropy can be calculated as
\begin{align*}
    \theta_{pt} =  [\log(1+\sigma^2(\Tilde{V}_{pt}^{MS}))] h(\Tilde{V}_{pt}^{MS})
\end{align*}
where $\Tilde{V}_{pt}^{MS}$ denotes MS coefficients of frame $t$ and patch $p$ passed through neural noise. There does not exist any frame rate bias since spatial entropies were calculated using only single frames and are frame rate agnostic. Spatial GREED (SGREED) is defined as :
\begin{align}
    \text{SGREED}_t = \frac{1}{P}\sum_{p=1} ^P |\theta_{pt} ^D - \theta_{pt} ^R|.
    \label{eqn:GSI}
\end{align}

\subsection{VMAF description}
VMAF uses multiple elementary quality indices and fuses them using a Support Vector Regressor (SVR) \cite{VMAF2016}. The first step in VMAF is extracting the following features : DLM \cite{li2011image}, VIF \cite{sheikh2006image} and absolute luma differences between consecutive frames (Temporal Information - TI). The DLM feature captures detail losses while VIF measures image fidelity based on Natural Scene Statistics (NSS). VIF is computed at four scales resulting four VIF features. TI aims to capture the temporal artifacts which are quantified through luma differences. This results in VMAF containing six features overall. TI is the only feature in VMAF that accounts for temporal quality evaluation.

\begin{algorithm}[t] \caption{GREED-VMAF Algorithm}
\begin{algorithmic}[1]
\Require reference video $R$, distorted video $D$
\Ensure GREED-VMAF features
\State Temporal subsample $R$ to get $PR$
\State Calculate DLM(PR,D)
\State Calculate VIF(PR,D) - 4 spatial scales
\State Scale $S = \{4,5\}$, band-pass filter bank $b_k: k \in \{1,\ldots 7\}$
\ForEach {$s \in S $}
\State Calculate SGREED from (\ref{eqn:GSI})
\ForEach {$b_k$}
\State Calculate TGREED$_k$ from (\ref{eqn:GTI})
\EndFor
\EndFor
\State Concatenate DLM, VIF, SGREED, TGREED to obtain GREED-VMAF features.
\end{algorithmic}
\label{alg1}
\end{algorithm}

\subsection{Feature Integration}
\label{sec:feat_integration}
Although GREED achieves high correlations against human judgements on LIVE-YT-HFR database, the performance is sensitive to the differences between the frame rates of reference and distorted sequences, where lower frame rate videos have lower correlations when compared to high frame rate videos. This behavior can be observed in Table \ref{Table:FPS_comparison}. Another shortcoming of GREED is that it underperforms in the presence of other types of distortions apart from frame rate and compression artifacts. This is evidenced in Table \ref{Table:VQA_database} where GREED is evaluated on databases such as LIVE-VQA \cite{seshadrinathan2010study}, CSIQ-VQA \cite{vu2014vis3} etc. VMAF also has a drawback in terms of capturing temporal distortions as TI feature is a basic measure and complex temporal artifacts such as ghosting, flickering etc. are not effectively represented by TI.

To overcome these limitations, we propose to combine the features from both GREED and VMAF to exploit the advantages offered by both models. Since GREED achieves superior performance for temporal artifacts, we replace the TI feature in VMAF by all GREED features. We retain the 5 spatial VMAF features (DLM and VIF from 4 scales) as they sufficiently capture spatial quality. GREED features are computed at 2 scales $s=4,5$ where entropic differences are computed after downscaling both spatial dimensions by $2^s$ times. For calculating TGREED Biorthogonal-2.2 band-pass filter with 3 levels of wavelet packet decomposition \cite{coifman1992entropy} was used. Since low-pass subband is ignored, this results in 7 TGREED features for each scale from respective temporal subbands and a total of 14 TGREED features from both scales. Although spatial components of VMAF account for spatial distortions, including SGREED features was experimentally observed to further improve performance. Thus the proposed model GREED-VMAF consists a total of 21 features altogether. The entire feature extraction procedure for GREED-VMAF is summarized in Algorithm \ref{alg1}.

\begin{table}[t]
\caption{Performance comparison of GREED against different FR algorithms on the LIVE-YT-HFR Database. The reported numbers are median values from 200 iterations of randomly chosen train-test sets. In each column first and second best models are boldfaced.}
    \label{Table:MOS_comparison}
    \centering
    \begin{tabular}{|c||c|c|c|c|}
        \hline
        ~    & SROCC $\uparrow$ & KROCC $\uparrow$ & PLCC $\uparrow$ & RMSE $\downarrow$ \\ \hline \hline
        PSNR & 0.7802 & 0.5934 & 0.7481 & 7.75 \\ 
        SSIM \cite{wang2004image} & 0.5566 & 0.4042 & 0.5418 & 9.99 \\ 
        MS-SSIM \cite{wang2003multiscale} & 0.5742 & 0.4135 & 0.5512 & 10.01 \\
        ST-RRED \cite{soundararajan2012video} & 0.6394 & 0.4516 & 0.6073 & 9.58 \\ 
        SpEED \cite{bampis2017speed} & 0.6051 & 0.4437 & 0.5206 & 10.28 \\
        VMAF \cite{VMAF2016}& 0.7782 & 0.5918 & 0.7419 & 8.10 \\ \hline
        GREED & \textbf{0.8822} & \textbf{0.7046} & \textbf{0.8869} & \textbf{5.48} \\
        GREED-VMAF & \textbf{0.8658} & \textbf{0.6840} & \textbf{0.8723} & \textbf{5.89} \\
        \hline
    \end{tabular}
\end{table}

\section{Experiments and Results}
\label{sec:experiments}
\subsection{Experimental Settings}
\textbf{Compared Methods} : We choose 3 FR-IQA methods: PSNR, SSIM \cite{wang2004image} and MS-SSIM \cite{wang2003multiscale}, and 3 FR-VQA models: ST-RRED \cite{soundararajan2012video}, SpEED \cite{bampis2017speed} and VMAF \cite{VMAF2016} for performance comparison. The above mentioned models require reference and distorted versions to have same frame rates, thus for the cases where the frame rates were different the distorted video was temporally upsampled to match the frame rate of reference video by frame duplication. Although the frame rates can be matched by temporally downsampling the reference video as well, we avoid this option as it can potentially introduce temporal artifacts in the reference video which is not desirable.

\textbf{Evaluation Criteria} : VQA models are compared in terms Spearman's rank order correlation coefficient (SROCC), Kendall's rank order correlation coefficient (KROCC), Pearson's linear correlation coefficient (PLCC), and root mean squared error (RMSE). PLCC and RMSE are calculated after passing the predicted scores through a four-parameter logistic non-linearity as described in \cite{VQEG2000}:
\begin{align}
    Q(x) = \beta_2 + \frac{\beta_1 - \beta_2}{1 + \exp\Bigg(-\Big(\frac{x - \beta_3}{|\beta_4|}\Big)\Bigg)}.
    \label{eqn:logistic_non}
\end{align}

\begin{table*}[t]
	\caption{Performance comparison of GREED against various FR methods for individual frame rates on the LIVE-YT-HFR Database. The reported numbers are median values over 200 iterations of randomly chosen train-test sets. In each column the best and second best values are marked in boldface.}
	\label{Table:FPS_comparison}
	\centering
	\scriptsize
	\scalebox{0.88}{
		\begin{tabular}{|c||c|c|c|c|c|c|c|c|c|c|c|c|c|c|}
			\hline
			& \multicolumn{2}{|c|}{24 fps} & \multicolumn{2}{|c|}{30 fps} & \multicolumn{2}{|c|}{60 fps} & \multicolumn{2}{|c|}{82 fps} & \multicolumn{2}{|c|}{98 fps} & \multicolumn{2}{|c|}{120 fps} & \multicolumn{2}{|c|}{Overall} \\
			\cline{2-15}
			~ & SROCC$\uparrow$ & PLCC$\uparrow$ & SROCC$\uparrow$ & PLCC$\uparrow$ & SROCC$\uparrow$ & PLCC$\uparrow$ & SROCC$\uparrow$ & PLCC$\uparrow$ & SROCC$\uparrow$ & PLCC$\uparrow$ & SROCC$\uparrow$ & PLCC$\uparrow$ & SROCC$\uparrow$ & PLCC$\uparrow$\\ \hline \hline
			PSNR & 0.541 & 0.487 & 0.564 & 0.535 & 0.753 & 0.694 & 0.771 & 0.771 & 0.821 & 0.783 & 0.741 & 0.736 & 0.780 & 0.748 \\ 
			SSIM \cite{wang2004image} & 0.266 & 0.222 & 0.283 & 0.189 & 0.382 & 0.302 & 0.371 & 0.362 & 0.537 & 0.497 & 0.867 & 0.833 & 0.556 & 0.541 \\ 
			MS-SSIM \cite{wang2003multiscale} & 0.305 & 0.260 & 0.296 & 0.238 & 0.416 & 0.338 & 0.439 & 0.393 & 0.578 & 0.561 & 0.706 & 0.696 & 0.574 & 0.551\\ 
			ST-RRED \cite{soundararajan2012video} & 0.305 & 0.275 & 0.296 & 0.206 & 0.612 & 0.613 & 0.584 & 0.513 & 0.650 & 0.604 & 0.755 & 0.696 & 0.639 & 0.607\\ 
			SpEED \cite{bampis2017speed} & 0.432 & 0.273 & 0.410 & 0.233 & 0.439 & 0.292 & 0.546 & 0.390 & 0.578 & 0.471 & 0.758 & 0.739 & 0.605 & 0.520\\
			VMAF \cite{VMAF2016} & 0.250 & 0.368 & 0.362 & 0.471 & 0.630 & 0.680 & 0.734 & 0.793 & \textbf{0.860} & 0.868 & 0.818 & 0.816 & 0.779 & 0.742\\  \hline
			GREED & \textbf{0.727} & \textbf{0.822} & \textbf{0.702} & \textbf{0.843} & \textbf{0.732} & \textbf{0.840} & \textbf{0.818} & \textbf{0.896} & \textbf{0.864} & \textbf{0.891} & \textbf{0.888} & \textbf{0.895} & \textbf{0.882} & \textbf{0.887}\\
			GREED-VMAF & \textbf{0.748} & \textbf{0.805} & \textbf{0.743} & \textbf{0.833} & \textbf{0.773} & \textbf{0.836} & \textbf{0.786} & \textbf{0.880} & \textbf{0.860} & \textbf{0.899} & \textbf{0.881} & \textbf{0.903} & \textbf{0.865} & \textbf{0.872}\\
			\hline
		\end{tabular}
	}
\end{table*}

\begin{table}[t]
\caption{SROCC performance comparison on multiple VQA databases. The reported numbers are median values from every possible combination of train-test splits with 80\% of content used for training. In each column the best and second best models are boldfaced. }
    \label{Table:VQA_database}
    \centering
    \footnotesize
    \begin{tabular}{|c||c|c|c|}
        \hline
        ~    & LIVE-VQA  & LIVE-mobile & CSIQ-VQA \\ \hline \hline 
        PSNR & 0.711 & 0.788 & 0.579 \\ 
        SSIM \cite{wang2004image} & 0.802 & 0.798 & 0.705 \\ 
        MS-SSIM \cite{wang2003multiscale} & \textbf{0.830} & 0.800 & 0.757 \\
        ST-RRED \cite{soundararajan2012video} & \textbf{0.826} & 0.882 & \textbf{0.813} \\ 
        SpEED \cite{bampis2017speed} & 0.801 & 0.886 & 0.743 \\ 
        VMAF \cite{VMAF2016}& 0.794 & \textbf{0.897} & 0.618 \\ \hline
        GREED & 0.750 & 0.863 & 0.780 \\
        GREED-VMAF & 0.784 & \textbf{0.904} & \textbf{0.907} \\
        \hline
    \end{tabular}
\end{table}

\subsection{Performance Analysis on LIVE-YT-HFR Dataset}
Since GREED-VMAF is a training based model, the LIVE-YT-HFR database is divided into 70\%, 15\% and 15\% subsets corresponding to training, validation and testing respectively. The validation subset was used to determine hyperparameters of SVR. Additionally we ensured that there was no content overlap between the subsets. To avoid any performance bias towards the choice of training set, the experiment was repeated 200 times with random train-test splits and median performance is reported. Since only the spatial features from VMAF are employed in GREED-VMAF, and since VMAF framework requires the compared videos to have same frame rate, we match the frame rate by temporally subsampling the reference video (PR video) instead of temporally upsampling distorted version as we observed the spatial artifacts were better captured by employing this design. SVR with linear kernel was used to learn mapping from features to quality scores.

In Table \ref{Table:MOS_comparison} the performance of GREED-VMAF against other models is compared. GREED and GREED-VMAF outperform other models by a large margin. In Table \ref{Table:FPS_comparison} the LIVE-YT-HFR database is subdivided into sets containing videos of same frame rate, and the performance is individually analyzed on each frame rate. Here as well GREED and GREED-VMAF have superior correlation when compared to other models across each frame rate. Although we observe a small drop in performance of GREED-VMAF in Table \ref{Table:MOS_comparison}, its performance is superior to GREED for individual frame rates as shown in Table \ref{Table:FPS_comparison}, particularly in low frame rate sets such as 24 and 30 fps. This makes performance of GREED-VMAF be less sensitive to the choice of distorted frame rate and provide more stable quality predictions than GREED.

\subsection{Performance Comparison on other VQA Databases}
In this subsection we evaluate the generalizability of GREED-VMAF features by comparing its performance on three popular VQA databases LIVE-VQA \cite{seshadrinathan2010study}, LIVE-mobile \cite{moorthy2012video} and CSIQ-VQA \cite{vu2014vis3}. These databases contain both the reference and distorted videos at the same frame rate, thus the TGREED term in expression (\ref{eqn:GTI}) will only depend on the absolute difference of scaled entropies since the ratio term simplifies to one. We divide the database into sets of 80\% and 20\% for training and testing respectively. We also ensured that there does not exist any overlap of contents between training and testing subsets. Further, to avoid any performance bias towards the choice of train-test contents, we repeat the above procedure for all possible train-test combinations, and the median SROCC performance is reported in Table \ref{Table:VQA_database}. For these datasets, SVR with radial basis function (RBF) kernel was used to learn mapping from features to quality scores. From the Table it maybe observed that GREED-VMAF outperforms GREED across all databases indicating that the combined set of features are more generalizable than GREED. Notably GREED-VMAF outperforms VMAF and achieves top performance among the compared models on LIVE-mobile and CSIQ-VQA datasets. Since GREED-VMAF performance is higher than both GREED and VMAF, it can be inferred that the respective individual features capture complementary perceptual quality information, and thus fusing them leads to better correlation with subjective judgments.

\section{Conclusion}
\label{sec:conclusion}
We proposed a simple extension to existing VQA models GREED and VMAF by combining the spatial features of VMAF with spatio-temporal features of GREED. This models exploits the advantages of both the models: can be used when reference and distorted videos have different frame rates, with no additional temporal pre-processing and can effectively capture other artifacts apart from frame rate and compression artifacts. We conducted a holistic evaluation of the proposed model on LIVE-YT-HFR database and observed that GREED-VMAF provides better correlation with perceptual judgments when videos of fixed frame rates were analyzed. We also evaluated the generalizability of GREED-VMAF on multiple fixed frame rate VQA databases and observed to have superior performance when compared to both GREED and VMAF. As part of future work we wish to further evaluate the generalizability of GREED-VMAF on additional VQA datasets such as NFLX, EPFL-Polimi etc. which is not reported in this work.
\bibliographystyle{IEEEtran}
\bibliography{ref}
\end{document}